\documentclass[twocolumn,pra,superscriptaddress]{revtex4-1}
\usepackage{eurosym}
\usepackage{amsmath,amssymb,amstext}
\usepackage{ulem}
\usepackage[usenames,dvipsnames]{color}
\usepackage{graphicx}
\usepackage{braket}
\usepackage{natbib}
\usepackage{comment}
\usepackage{dcolumn}
\usepackage[english]{babel}
\usepackage{wasysym}
\usepackage{bm}
\usepackage[colorlinks,bookmarks=false,citecolor=blue,linkcolor=red,urlcolor=blue]{hyperref}

\usepackage{appendix}
\usepackage{textcomp,booktabs}
\usepackage[usenames,dvipsnames]{color}
\usepackage{mathrsfs}
\usepackage{float}
\usepackage{xcolor}
\usepackage{colortbl}
\usepackage{amsmath,amssymb,enumerate,epsfig,bbm,calc,color,ifthen,capt-of}
\makeatletter

\def\CT@@do@color{%
	\global\let\CT@do@color\relax
	\@tempdima\wd\z@
	\advance\@tempdima\@tempdimb
	\advance\@tempdima\@tempdimc
	\advance\@tempdimb\tabcolsep
	\advance\@tempdimc\tabcolsep
	\advance\@tempdima2\tabcolsep
	\kern-\@tempdimb
	\leaders\vrule
	\hskip\@tempdima\@plus  1fill
	\kern-\@tempdimc
	\hskip-\wd\z@ \@plus -1fill }
\makeatother

\newcommand{\beq}{\begin{equation}}
\newcommand{\eeq}{\end{equation}}
\newcommand{\beqa}{\begin{eqnarray}}
\newcommand{\eeqa}{\end{eqnarray}}

\def\jpb#1{{ J.\ Phys.\ B} {\bf#1}}

\def\pra#1{{ Phys.\ Rev. A\/} {\bf#1}}
\def\prb#1{{ Phys.\ Rev. B\/} {\bf#1}}
\def\pre#1{{ Phys.\ Rev. E\/} {\bf#1}}
\def\prl#1{{ Phys.\ Rev.\ Lett.} {\bf#1}}
\def\pr#1{{ Phys.\ Rev.\/} {\bf#1}}

\def\rmp#1{{ Rev.\ Mod.\ Phys.} {\bf#1}}

\begin{document}

\title{Extended virtual detector theory including quantum interferences}
 
\author{Rui-Hua Xu}
\affiliation{Institute of Applied Physics and Computational Mathematics, Beijing 100094, China}

\author{Xu Wang}
\email{xwang@gscaep.ac.cn}
\affiliation{Graduate School, China Academy of Engineering Physics,  Beijing 100193, China}

\author{J. H. Eberly}
\affiliation{Department of Physics and Astronomy, University of Rochester, Rochester, NY 14627, USA}

\date{\today}

\begin{abstract}

We extend our earlier ``virtual detector" method [X. Wang, J. Tian, and J. H. Eberly, Phys. Rev. Lett. {\bf 110}, 243001 (2013)], a hybrid quantum mechanical and classical trajectory method, to include phases in the classical trajectories. Effects of quantum interferences, lost in the earlier method, are restored. The obtained photoelectron momentum distributions agree well with the corresponding numerical solutions of the time-dependent Schr\"odinger equation.
	
\end{abstract}

\maketitle
\section{Introduction}

The interaction between isolated atoms or molecules with intense laser fields has led to highly nonlinear phenomena such as above-threshold ionization \cite{Agostini1979, Yang1993,Paulus1994, Lewenstein1995}, high harmonic generation \cite{McPherson1987, Ferray1988, Lewenstein1994}, nonsequential double ionization \cite{Walker1994, Palaniyappan2005, Becker2012}, attosecond physics \cite{Krausz2009, Pazourek2015}, etc. Theoretical descriptions are not trivial due to the nonperturbative nature of such an interaction. Various approaches have been developed, each having some particular advantages and of course also limitations. 

The most accurate approach would be numerically solving the time-dependent Schr\"odinger equation (TDSE) \cite{Tong1997, Parker2001, Parker2006, Schneider2006, Morishita2007, Hu2010}. The disadvantages of this approach include high computational loads and difficulties in providing clear physical pictures.  The theory of strong-field approximation is analytically simple and computationally efficient \cite{Keldysh1965, Faisal1973, Reiss1980}. Qualitative agreements with experiments are usually obtained due to approximating the continuum states with Volkov states and neglecting the atomic excited states.

Classical simulations are widely used to gain insights into strong-field processes, especially double or multiple ionization processes \cite{Panfili2001, Ho2005, Mauger2009, Wang2009, Chaloupka2016}, which are extremely difficult for quantum mechanical calculations. The advantages include clear physical pictures and very moderate computational loads. The disadvantage is that only qualitative agreements can be expected with experiments.

Classical models that incorporate some quantum mechanical features have also been developed \cite{Brabec1996, Chen2000, Ye2008}, with the expectation that better agreements with experiments can be obtained. In these models, the electrons obey classical mechanics, but their initial conditions (positions, momenta, weights) are assigned according to a quantum tunneling picture, exploiting e.g. the Ammosov-Delone-Krainov tunneling formula \cite{ADK}. These models have been further developed to include phases accumulated along the classical trajectories, such that features of quantum interferences are at least partially retained \cite{Li2014, Geng2014, Shvetsov2016}.

It is to be pointed out that the above semi-classical models rely on an assumed (adiabatic) tunneling-ionization mechanism, which is an approximation to the real, much more complicated, ionization process in the quasistatic limit (characterized by a Keldysh parameter $\gamma \ll 1$ \cite{Keldysh1965}). In fact, most strong-field experiments are not performed in that limit, but with $\gamma \sim 1$. The models may be modified \cite{Song2016} by using the (nonadiabatic) Perelomov-Popov-Terenr'ev ionization formula \cite{PPT}. Debates have been reported in the literature on how to specify the initial conditions of the electron trajectories starting from the tunneling exit. Where is exactly the tunneling-exit point? (See, e.g., a summary of several different criteria in \cite{Ni2018}) What should be the velocity of the electron at the tunneling exit point? \cite{Comtois2005, Pfeiffer2012, Sun2014, Camus2017, Tian2017, Xu2018}  It is found that the final electron momentum distribution depends sensitively on how these initial conditions are assigned.

We were motivated to develop a trajectory-based method that does not rely on ad hoc ionization mechanisms \cite{Wang2013}. The idea of our method is that the TDSE is first solved on a small numerical grid around the atom without pre-assuming an ionization mechanism, and then the outgoing wave is transformed into classical trajectories. The expectation is that this method combines the advantages of both the accuracy of quantum mechanics and the computational efficiency of classical mechanics.

The way to do this wave-to-trajectory transformation is to use the concept of probability current. The velocity of the probability current at some position is used as the initial velocity of a classical trajectory. A nickname of ``virtual detector" (VD, virtually detecting the velocity of the wave at some position) was given to such a method by Feuerstein and Thumm \cite{Feuerstein2003}, but we comment that using the velocity of the probability current to initiate a trajectory was exactly the proposal of Bohm \cite{Bohm-1, Bohm-2} and has been widely used in other fields as well \cite{Wyatt2005}. 

The VD method has been shown to yield good agreements with TDSE calculations on photoelectron momentum distributions \cite{Wang2013, Feuerstein2003, Wang2018}, although quantum interferences are necessarily lost due to the transformation to classical trajectories. Besides, the VD technique has been found very useful in gaining insights into the ultrafast ionization process \cite{Ni2016, Teeny20161, Teeny20162, Tian2017, Xu2018}.  
The goal of the current paper is to present a further development of the VD method. We extend our earlier method \cite{Wang2013} to include phases accumulated along the classical trajectories, therefore the effects of quantum interferences are retained, at least partially. Indeed, we show that discrete rings in the electron momentum distributions due to absorption of different number of photons are restored. 

This paper will be organized as follows. In Section II a brief introduction to the VD method will be given, and the way to include phases along classical trajectories will be explained. Numerical results and discussions will be given in Section III. A conclusion will be given in Section IV.

\section{Method}

\subsection{A brief introduction to the VD method}

We briefly summarize the VD method here, since more details have been given previously in \cite{Wang2013, Wang2018}. The space (centered at the atomic ion core) is separated into a relatively small inner region and an outer region. In the inner region, the TDSE is solved numerically. In the outer region, the electron is treated as a classical particle and its trajectory evolves according to classical mechanics. The key ingredient of this hybrid method is a VD network that encircles the atomic center, ``detecting" (but not affecting, hence virtual) the outgoing wave and transforming the wave into classical trajectories. Each trajectory is born from one of the VDs, with an initial velocity and also a weight assigned by the VD. Subsequent evolution of the trajectory till the end of the laser pulse is deterministic. Observables such as the final electron momentum distribution can be obtained by summing over all trajectories with weights.

The electron initially is assumed to be in the ground state of the atom. The VDs are placed evenly along a circle where the ionization process can be regarded as completed, e.g. 20 or 30 a.u. from the atomic ion. After the laser pulse is turned on, the electron wavefunction evolves according to the TDSE, and the VDs are triggered to initiate classical trajectories. Consider a classical trajectory initiated at time $t_0$ from a VD located at $\vec{r}_0$. The probability current is (atomic units are used unless otherwise stated)
 \beqa 
 \vec{j}(\vec{r}_0,t_0)
 &=&\frac{i}{2} \left[ \Psi(\vec{r}_0,t_0) \nabla \Psi^{*}(\vec{r}_0,t_0)-c.c. \right] \nonumber \\
 &=&A(\vec{r}_0,t_0)^2\nabla\phi(\vec{r}_0,t_0) \label{e.flux}
 \eeqa
where $A(\vec{r}_0,t_0)$ and $\phi(\vec{r}_0,t_0)$ are the amplitude and the phase of the wavefunction $\Psi(\vec{r}_0,t_0)$, respectively, both being real functions. 

The weight of the classical trajectory initiated at $\{ \vec{r}_0, t_0\}$ is given by the magnitude of the probability current \cite{Wang2018}
\beq
w(\vec{r}_0,t_0) = j(\vec{r}_0,t_0), \label{e.weight}
\eeq
and the initial momentum of the classical trajectory is given by the phase gradient
\beq
\vec{p}_0(\vec{r}_0,t_0)=\nabla\phi(\vec{r}_0,t_0). \label{e.mtm}
\eeq

The position and momentum of the electron at the end of pulse, $\{t_1, \vec{r}_1, \vec{p}_1\}$, can be determined from $\{t_0, \vec{r}_0, \vec{p}_0\}$ according to classical mechanics. After the laser pulse is over, the momentum of the electron no longer changes, although the electron will need time to fly from $\vec{r}_1$ to the detector located at $\vec{r}_D$. The final state of the electron is noted as $\{t_f, \vec{r}_D, \vec{p}_1\}$. The following relation holds for the free flying
\beq
\vec{r}_D - \vec{r}_1 = (t_f - t_1)\vec{p}_1. 
\eeq

The final momentum distribution is obtained by summing over all classical trajectories with final momenta around a given value $\vec{p}$
\beq
W(\vec{p}) = \sum_{\{\vec{r}_0,t_0\}} w(\vec{r}_0,t_0) \ \text{with} \ \vec{p}_1 \in \left[\vec{p}, \vec{p}+d\vec{p} \right]. \label{e.Sum1}
\eeq

\subsection{Phases accumulated along a classical trajectory}

\begin{figure}[t!]
  \centering
  \includegraphics[width=7cm]{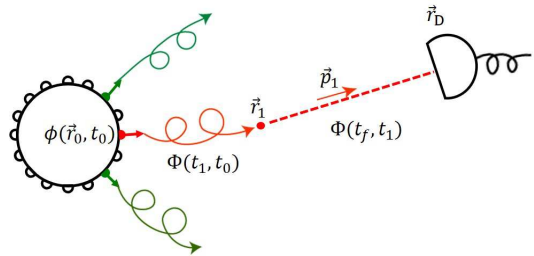}\\
   \caption{An illustration of the phases accumulated along a classical trajectory, from a virtual detector where it is initiated to a real detector where it is ended. The trajectory is initiated at time $t_0$ and position $\vec{r}_0$, where it obtains an initial phase $\phi(\vec{r}_0,t_0)$. It travels to $\vec{r}_1$ with the laser pulse which ends at time $t_1$. During this time the phase gained is $\Phi(t_1,t_0)$. Then it flies freely to the detector located at $\vec{r}_D$ with a constant momentum $\vec{p}_1$. The phase gained during the free flying is $\Phi(t_f,t_1)$.}\label{f.illustration}
\end{figure}

\begin{figure*}[t!]
  \centering
  \includegraphics[width=1.8\columnwidth]{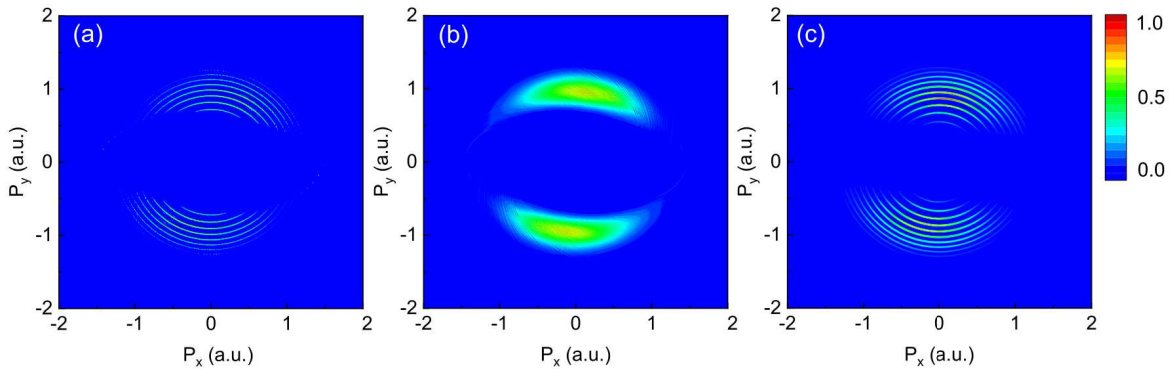}\\
   \caption{Electron momentum distributions obtained from (a) the extended VD method with phases, (b) the original VD method without phases, and (c) the numerical TDSE calculations under the same conditions. The laser wavelength is 600 nm and the intensity is 0.4 $\mathrm{PW/cm^2}$. }\label{f.600nm_full}
\end{figure*}

Here we explain how the phase should be taken into account. An illustration is given in Fig. \ref{f.illustration} along with the explanation. Consider also the classical trajectory initiated at $\{\vec{r}_0,t_0\}$, when the phase of the wavefunction is $\phi(\vec{r}_0,t_0)$. We take this phase as the initial phase of the trajectory. The trajectory then propagates from $t_0$ to $t_1$ in the laser field, and the phase accumulated is the time integral of the action along the trajectory 
\beq
\Phi(t_1,t_0) = \int_{t_{0}}^{t_{1}}L[\vec{p}(t),\vec{r}(t)]dt,
\eeq
where the Lagrangian is given by
\beq
L[\vec{p}(t),\vec{r}(t)]=\frac{p^{2}(t)}{2}-V[r(t)]-\vec{E}(t)\cdot\vec{r}(t).
\eeq
Here $V(r)$ is the potential from the atomic ion core and $\vec{E}(t)$ is the laser electric field. 

At time $t_1$ the laser pulse is over and the electron becomes a free particle. It will fly freely from $\vec{r}_1$ to the detector at $\vec{r}_D$ with a constant momentum $\vec{p}_1$, and the electron will gain an additional phase
\beq
\Phi(t_f, t_1) = \vec{p}_1 \cdot (\vec{r}_D-\vec{r}_1).
\eeq
Because only trajectories with the same final momentum interfere, the phase $\vec{p}_1\cdot \vec{r}_D$ is an overall phase and can be neglected.

The total phase associated with the trajectory is
\beqa
\Phi_{tot}(\vec{r}_0,t_0) &=& \phi(\vec{r}_0,t_0) + \Phi(t_1,t_0) +  \Phi(t_f, t_1) \nonumber\\
&=& \phi(\vec{r}_0,t_0) + \int_{t_{0}}^{t_{1}}L[\vec{p}(t),\vec{r}(t)]dt - \vec{p}_1\cdot \vec{r}_1. \nonumber\\
&&
\eeqa
The final momentum distribution will be obtained by summing over classical trajectories with final momenta around $\vec{p}$ ``coherently"
\beq
W(\vec{p}) = \left| \sum_{\{\vec{r}_0,t_0\}}  \sqrt{w(\vec{r}_0,t_0)} e^{i\Phi_{tot}(\vec{r}_0,t_0)} \text{with} \ \vec{p}_1 \in \left[\vec{p}, \vec{p}+d\vec{p} \right] \right|^2. \label{e.Sum2}
\eeq

\section{Results and discussions}

\subsection{Restoring the interference patterns}

Figure \ref{f.600nm_full} shows the electron momentum distributions of (a) the extended VD method with phases, (b) the original VD method without phases, and (c) numerical TDSE calculations under the same conditions. Indeed, one sees that the interference structures (i.e., discrete photon rings), which are lost in the original VD method, are restored by using the extended VD method. 

The laser intensity is 0.4 $\mathrm{PW/cm^2}$ and the wavelength is 600 nm. An elliptically  polarized laser pulse of the following form is used
\beq
\vec{E}(t)=E_{0}f(t) \left( \hat{x}\sin\omega t+\hat{y}\epsilon \cos\omega t \right),
\eeq
where $\epsilon=0.78$ is the ellipticity value, $\omega$ is the angular frequency, and $f(t)$ is a pulse envelope function. A trapezoidal pulse with a total duration of 20 optical cycles has been used, with a two-cycle linear turning on and a two-cycle linear turning off.

\begin{figure}[b!]
  \centering
  \includegraphics[width=8cm]{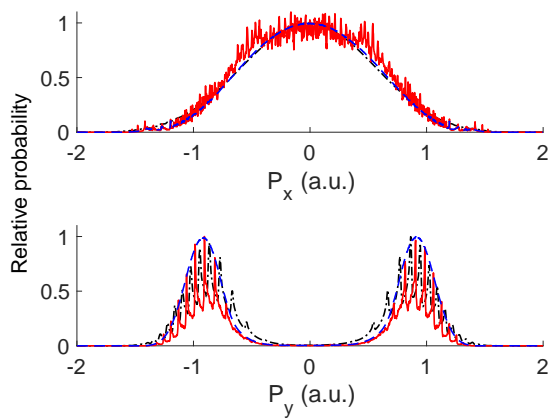}\\
   \caption{Projections of the electron momentum distributions shown in Fig. \ref{f.600nm_full} onto the \textrm{$P_{x}$} axis and the \textrm{$P_{y}$} axis. The red solid lines are for the extended VD method. The blue dashed lines are for the original VD method. The black dash-dotted lines are for the TDSE results.
 }\label{f.600nm_pxpy}
\end{figure}

The TDSE is numerically integrated on a two-dimensional grid, which is large enough (from -600 to 600 a.u. for each dimension) to keep all the population on the grid during the pulse. The ion core potential is given by $V(r)=-1/\sqrt{r^2+0.28}$, which yields a ground state energy of -0.9 a.u., matching that of the helium atom. The TDSE is integrated numerically using a split-operator method starting from the ground state, which is found using an imaginary-time technique.

The VD method uses a much smaller numerical grid (from -50 to 50 a.u. for each dimension). 400 VDs are placed evenly on a circle of radius 20 a.u. The VDs are triggered at each time step ($\Delta t = 0.01$ a.u.) as the wavefunction propagates out. An absorption region is used outside the VD circle to avoid wave reflection from the boundary of the small numerical grid.

We further compare the projections of the electron momentum distributions onto the $P_x$ axis and the $P_y$ axis, as shown in Fig. \ref{f.600nm_pxpy}. For the $P_x$ distribution, the three results agree well, although the results from the extended VD method seem somewhat noisy. The noisy structure can be gradually suppressed by using more trajectories, although the suppression may progress slowly \cite{Shvetsov2016, Song2016}. For the $P_y$ distribution, the extended VD method restores the interference peaks which are lost in the original VD method. Compared with the TDSE results, the extended VD method shows the same number of peaks, although the positions of the peaks show a small overall shift.

\begin{figure}[t!]
  \centering
  \includegraphics[width=8.4cm]{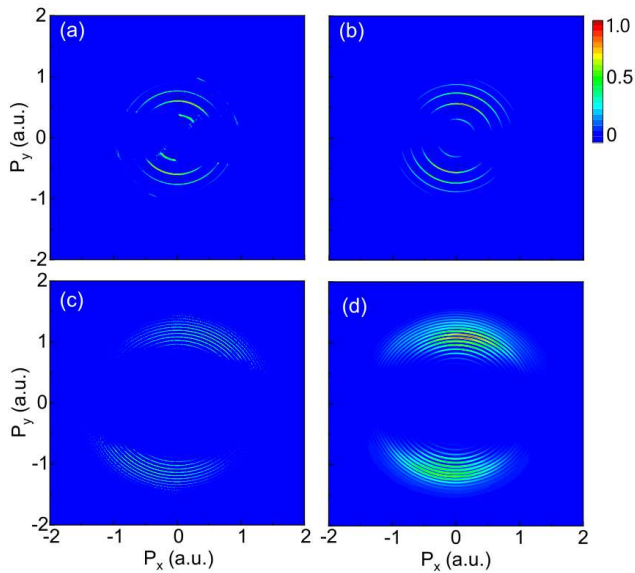}
  \caption{ Electron momentum distributions obtained from the extended VD method (left column) and from TDSE calculations (right column) with laser wavelength 400 nm (top row) and 740 nm (bottom row). The laser intensity is 0.4 $\mathrm{PW/cm^2}$ for both cases. }\label{f.400_740nm}
\end{figure}

In Fig. \ref{f.400_740nm}, we also plot electron momentum distributions under two different laser wavelengths, namely, 400 nm and 740 nm. These figures also present clear interference patterns as the TDSE results show.

\subsection{Some critical remarks}

{\it On semi-classical trajectory methods in general.} The area of semi-classical physics is not a concluded one. There is no clear border between the quantum world and the classical world, and there is no single perfect approach to reconcile the wave description and the trajectory description. It is therefore advisable to keep an open mind as well as a critical attitude to any semi-classical trajectory method, acknowledging its advantages, disadvantages, and domain of applicability. 

{\it On the probability current.} As mentioned in the Introduction, using the velocity of the probability current to initiate a trajectory is a widely used technique in different areas of theoretical physics or chemistry \cite{Bohm-1, Bohm-2, Wyatt2005}. The concept of the probability current, however, has its limitations that are not widely appreciated. An obvious fact is that the probability current is zero everywhere for a bound state with magnetic quantum number $m=0$. A trajectory approach based on the velocity of the probability current would say that the electron is static in these bound states. This sounds odd, at least apparently. Efforts, however, have been seen in the literature attempting to extend the standard concept of the probability current, e.g. by the group of E. J. Heller at Harvard University \cite{Mason2012, Mason2013, Mason2015}.

We also point out that the velocity of the probability current at some position, $\nabla \phi(\vec{r},t)$, is actually the average value of all possible velocities at that position \cite{Wang2018}, if we understand it from the phase-space perspective using, e.g., the Wigner function. This is actually the main source of error for the VD method, probably for all trajectory methods based on the concept of probability current. Representing a distribution with a single average value has of course only limited domain of applicability.

\section{Conclusion}

In this paper we present an extension to our previous virtual detector method \cite{Wang2013}, which is a hybrid quantum mechanical and classical trajectory method and is expected to combine some advantages from both sides. The extension made here is to include phases in the classical trajectories, so that effects of quantum interferences are retained, at least partially.

We explain how the phases should be included using a simple physical picture as given in Section II. The numerical results indeed show that the lost quantum interferences are restored in the current extended method. The electron momentum distributions show interference structures as features of absorbing different number of photons, agreeing with the results of TDSE calculations under the same condition.

We acknowledge support from NSFC No. 11947204 and 11774323, NSAF No. U1930403, and US DOE Grant No. DE-FG02-05ER15713.

\end{document}